\documentclass[twocolumn]{aastex631}
\usepackage{amsmath}

\newcommand{\EDE}{\mathrm{EDE}}
\newcommand{\LCDM}{$\Lambda$CDM }

\shorttitle{New constraint on Early Dark Energy using the profile likelihood}
\shortauthors{Herold et al.}

\begin{document}

\title{New constraint on Early Dark Energy from \textit{Planck} and BOSS data using the profile likelihood}

\author[0000-0001-9054-1414]{Laura Herold}
\correspondingauthor{Laura Herold}
\affiliation{Max-Planck-Institut für Astrophysik, Karl-Schwarzschild-Str. 1, 85748 Garching, Germany}
\email{lherold@mpa-garching.mpg.de}

\author[0000-0002-5032-8368]{Elisa G. M. Ferreira}
\affiliation{Max-Planck-Institut für Astrophysik, Karl-Schwarzschild-Str. 1, 85748 Garching, Germany}
\affiliation{Kavli IPMU (WPI), UTIAS, The University of Tokyo, 5-1-5 Kashiwanoha, Kashiwa, Chiba 277-8583, Japan}
\affiliation{Instituto de F\'isica, Universidade de S\~ao Paulo - C.P. 66318, CEP: 05315-970, S\~ao Paulo, Brazil}
\email{elisa.ferreira@ipmu.jp}

\author[0000-0002-0136-2404]{Eiichiro Komatsu}
\affiliation{Max-Planck-Institut für Astrophysik, Karl-Schwarzschild-Str. 1, 85748 Garching, Germany}
\affiliation{Kavli IPMU (WPI), UTIAS, The University of Tokyo, 5-1-5 Kashiwanoha, Kashiwa, Chiba 277-8583, Japan}
%\email{komatsu@mpa-garching.mpg.de}

\begin{abstract}
       A dark energy-like component in the early universe, known as early dark energy (EDE), is a proposed solution to the Hubble tension. Currently, there is no consensus in the literature as to whether EDE can simultaneously solve the Hubble tension and provide an adequate fit to the data from the cosmic microwave background (CMB) and large-scale structure of the universe. 
       In this work, we deconstruct the current constraints from the \textit{Planck} CMB and the full-shape clustering data of the Baryon Oscillation Spectroscopic Survey (BOSS) to understand the origin of different conclusions in the literature. We use two different analyses, a grid sampling and a profile likelihood, to investigate whether the current constraints suffer from volume effects upon marginalization and are biased towards some values of the EDE fraction, $f_\mathrm{EDE}$. We 
       find that the $f_\mathrm{EDE}$ allowed by the data strongly depends on the particular choice of the other parameters of the model and that several choices of these parameters prefer larger values of $f_\mathrm{EDE}$ than in the Markov Chain Monte Carlo analysis. This suggests that volume effects are the reason behind the disagreement in the literature. Motivated by this, we use a profile likelihood to analyze the EDE model and compute a confidence interval for $f_\mathrm{EDE}$, finding $f_\mathrm{EDE} = 0.072\pm 0.036$ ($68\%$~C.L.). Our approach gives a confidence interval that is not subject to volume effects and provides a powerful tool to understand whether EDE is a possible solution to the Hubble tension. 
\end{abstract}

%%%%%%%%%%%%%%%%% BODY OF PAPER %%%%%%%%%%%%%%%%%%

\section{Introduction}
\label{sec:Intro}

% HUBBLE TENSION
Measurements of the Hubble constant, $H_0$, the present-day expansion rate of the universe, obtained with different techniques show a discrepancy known as the ``Hubble tension'' \citep{Bernal_2016}. Indirect measurements, which depend on the assumption of a cosmological model, yield systematically lower values of $H_0$ than direct measurements, which do not or only weakly depend on the assumption of a cosmological model. 

The most significant tension is seen between the (indirect) inference of $H_0$ from the cosmic microwave background (CMB) data of the \textit{Planck} mission assuming a flat $\Lambda$ Cold Dark Matter ($\Lambda$CDM) cosmological model, $H_0 = 67.37\pm 0.54\,$km/s/Mpc \citep{Planck_col_2020}, and the (direct) local inference from Cepheid-calibrated Type Ia supernovae of the SH0ES project, $H_0 = 73.04 \pm 1.04 \,$km/s/Mpc \citep{Riess:2021jrx}. The statistical significance of the tension is $5\sigma$. Throughout this paper, we quote uncertainties at the 68\,\% confidence level (C.L.), unless noted otherwise.

% EDE
This tension could hint at new physics beyond the flat $\Lambda$CDM model. One of the proposed models to alleviate the tension is early dark energy (EDE) \citep{Poulin_2018, Poulin_2019, Smith_2020}. In this model, the $\Lambda$CDM cosmology is extended to include a dark energy-like component in the pre-recombination era, which reduces the size of the sound horizon and increases $H_0$ \citep{Bernal_2016}. EDE is typically parametrized by three parameters: the initial value of the EDE field ($\theta_i$), its maximum fractional energy density ($f_\EDE$) and the critical redshift ($z_c$) at which this maximum fraction is reached. 

% STATUS OF THE FIELD
EDE was shown to reduce the tension between the values of $H_0$ \citep{Poulin_2018, Smith_2020} inferred from the CMB data of \textit{Planck} \citep{Planck_col_2016}, the baryon acoustic oscillation (BAO) and the redshift-space distortion data of the Baryon Oscillation Spectroscopic Survey \citep[BOSS;][]{BOSS_col_2017}, the BAO measurements from the 6-degree Field Galaxy Survey \citep[6dFGS;][]{Beutler_2011} and Sloan Digital Sky Survey Main Galaxy Sample \citep[SDSS MGS;][]{Ross_2015}, the Pantheon supernova sample \citep{Scolnic_2018}, and the direct measurement by the SH0ES collaboration \citep{Riess_2019}. They find %a maximum fractional energy density
$f_\EDE=0.107^{+0.035}_{-0.030}$, which gives %a value of 
$H_0 =71.49\pm 1.20\,$km/s/Mpc.

However, it was pointed out in \citet{Hill_2020} that introducing EDE leads to a higher amplitude of matter density fluctuations parametrized by $\Omega_m$ and $\sigma_8$, worsening the so-called $\sigma_8$-tension.
They showed that including further large-scale structure (LSS) probes such as Dark Energy Survey \citep[DES;][]{Abbott_2018}, Kilo-Degree Survey \citep[KiDS-VIKING;][]{Hildebrandt_2020} and Hyper Suprime-Cam \citep[HSC;][]{Hikage_2019}, which are particularly sensitive to  $\Omega_m$ and $\sigma_8$, weakens the evidence for EDE. When including all probes but $H_0$ from SH0ES, their analysis yields an upper limit of $f_\EDE < 0.06$  at 95\%~C.L. A similar constraint of $f_\EDE <0.072$ at 95\%~C.L (with $f_\EDE = 0.025_{-0.025}^{+0.006}$) is obtained when employing the full shape of the galaxy power spectrum combined with the BAO data of BOSS Data Release 12 (DR 12) galaxies along with the \textit{Planck} data \citep{Ivanov_2020}.  Concurrently, a similar analysis from \citet{DAmico_2021} found $f_\EDE < 0.08$ at 95\%~C.L for the same data set and additionally including the Pantheon supernova sample. These three papers conclude that EDE does not solve the Hubble tension.  

In the analyses of \citet{Hill_2020}, \citet{Ivanov_2020}, and \citet{DAmico_2021}, all three EDE parameters $\left\{f_\EDE,\theta_i, z_c\right\}$ are varied, which is referred to as the ``3-parameter model.''
%Shortly after, 
\citet{Smith_2021} argued that the reason for the small preferred value of $f_\EDE$ found by them is due to volume effects upon marginalization, and proposed alternative approaches.\footnote{An exploration of volume effects with an averaging method can be found in the appendix of \citet{Ivanov_2020}.}
In particular, they found $f_\EDE = 0.072\pm0.034$ for the same data set as in \cite{Ivanov_2020}, when fixing two EDE parameters $\left\{ \theta_i, z_c \right\}$, which is referred to as the ``1-parameter model.'' Within the 1-parameter model, they observe that including LSS data decreases the evidence for EDE similar to the 3-parameter model; they relate this tighter constraint on EDE to the lower clustering amplitude preferred by LSS data compared to CMB data.
The 1-parameter model was already explored earlier in \cite{Niedermann_2020} in the context of new EDE.

Currently there is no agreement in the community as to whether EDE can simultaneously solve the Hubble tension and fit all available data sets. A new chapter in this discussion was presented recently: Two groups \citep{Hill_2021_ACT, Poulin_2021_ACT} reported independently on a $2-3\sigma$ preference for EDE when analyzing the model using the CMB data of the Atacama Cosmology Telescope \citep[ACT;][]{Choi_2020}. South Pole Telescope data \citep{Dutcher_2021} is consistent with both ACT and \textit{Planck} results \citep{LaPosta_2021}.

% WHAT ARE THE PROBLEMS
One question that remains open is: What is the reason behind this disagreement?
The root of this seems to lie in the Markov Chain Monte Carlo (MCMC) sampling of the three parameters of the EDE model. For $f_\EDE =0$, the EDE model is degenerate with $\Lambda$CDM for any choice of $\theta_i$ and $z_c$. Therefore, the parameter volume for $f_\EDE =0$ is larger than for every $f_\EDE >0$. This can lead to a preference for $f_\EDE=0$ in the marginalized posterior, affecting the inferred amount of EDE allowed by the data. On the other hand, fixing some parameters of the model, as for the 1-parameter model, is an incomplete analysis, as stated in \citet{Smith_2021}; the results might depend on the particular choice of the parameters. %, and might be strongly dependent on the values of the fixed parameters. Therefore, this analysis cannot be used to reach a definitive answer to the puzzle. 

% DECONSTRUCTION AND SOLUTION
In this paper, we deconstruct the current constraints on the EDE model from the CMB and BOSS full-shape clustering data. Our goal is to understand where the disagreement in the literature comes from and to check if volume effects are indeed present. In particular, we answer the following questions: Is the 3-parameter model affected by the two unconstrained parameters $\theta_i$ and $z_c$ or by volume effects? Do the results of the 1-parameter model depend on the particular choice of $\theta_i$ and $z_c$ and how well can the results be generalized to the full 3-parameter model? How would the constraints on $f_\EDE$ change if those effects were eliminated?

To this end, we perform two analyses: a grid sampling and a profile likelihood. With the grid sampling, we explore the parameter space of $\left\{\theta_i, z_c\right\}$ by fixing them to a wide range of values and performing the 1-parameter analysis. This analysis shows that higher values of $f_\EDE$ are consistent with the data, which suggests that
the 3-parameter MCMC analysis is affected by volume effects, and that there is a strong dependence of $f_\EDE$ on the particular choice of $\left\{\theta_i, z_c\right\}$. This makes it difficult to generalize the results of the 1-parameter model. To confirm the presence of volume effects, we perform a frequentist-statistic analysis using a profile likelihood. We find that a considerably larger $f_\EDE$ is preferred by the data compared to the Bayesian MCMC analysis, confirming that volume effects affect the 3-parameter analysis. 

The rest of this paper is organized as follows. In Section~\ref{sec:Model}, we describe the EDE model. In Section~\ref{sec:Deconstructing}, we deconstruct the current constraints using the grid and the profile likelihood. In Section~\ref{sec:Constructing}, we construct a new confidence interval using the profile likelihood. We discuss the results and conclude in Section~\ref{sec:Discussion}.

%%%%%%%%%%%%%%%%%%%%%%%%%%%%%%%%%%%

\section{EDE Model}
\label{sec:Model}

The idea behind early-time solutions to the Hubble tension is to reduce the sound horizon and hence increase the inferred value of $H_0$ \citep{Bernal_2016}. The sound horizon, $r_s = \int_{z_{*}}^{\infty} c_s(z) \, dz / H(z) $, where $z_{*}$ is the redshift of the last scattering surface, $c_s(z)$ the sound speed in the baryon-photon plasma, and $H(z)$ the expansion rate of the universe, is dominated by contributions near the lower bound of the integral. 

EDE \citep{kamionkowski_2014,Karwal:2016vyq,Caldwell_2018} is an extra component added to the energy density budget near $z_*$, which increases $H(z)$ and lowers $r_s$.
This can be achieved by a pseudo scalar field, $\phi$, which obeys the following requirements: (i) it starts becoming relevant at matter-radiation equality; (ii) it behaves like dark energy at early times; and (iii) its energy density dilutes faster than the matter density after $z_*$. To model this behavior, the canonical EDE model is given by the potential \citep{Poulin_2019}:
\begin{equation}
\label{eq:potential}
    V(\phi) = V_0 \, \left[ 1 - \cos (\phi/f) \right]^n \,,
\end{equation}
where $V_0 = m^2 f^2$, $m$ is the mass of $\phi$, and $f$ is the spontaneous symmetry breaking scale.  

The parameters of the model can be re-written in terms of the phenomenological parameters $\left\{ f_\EDE, \theta_i,z_c, n \right\}$, where $f_\EDE$ is the maximum fraction of EDE at the critical redshift $z_c$, and $\theta_i$ is the initial value of the dimensionless field, $\theta\equiv \phi/f$. A larger value of $f_\EDE$ leads to a higher $H_0$. To solve the Hubble tension, it was predicted that $f_\EDE\simeq 0.1$ would be necessary \citep{Knox:2019rjx}.

The EDE field $\phi$ in a cosmological background with the potential given in Equation~(\ref{eq:potential}) behaves like dark energy initially, with the field essentially frozen. Once $H(z)$ becomes smaller than the effective mass $m_{\mathrm{eff}} = d^2 V(\phi)/d\phi^2$, $\phi$ starts decaying and oscillating at the bottom of the potential with an effective, time-averaged equation of state parameter of $\langle w \rangle = (n-1)/(n+1)$. Here, we choose $n=3$ as in the previous analyses, which was shown to dilute sufficiently fast to satisfy the requirement (iii) \citep{Poulin_2019,Smith_2020}.

%%%%%%%%%%%%%%%%%%%%%%%%%%%%%%%%%%%
\vspace{1.5cm}
\section{Deconstructing the current constraints on the EDE model}
\label{sec:Deconstructing}

\subsection{Data and methodology}
\label{sec:Method}

For our analysis, we use a similar setup as in \citet{Ivanov_2020}. We combine the following publicly available extensions of the Einstein--Boltzmann solver \texttt{CLASS} \citep{Lesgourgues_2011, Blas_2011}:\footnote{The code used for this analysis is publicly available at \\\url{https://github.com/LauraHerold/CLASS-PT_EDE}.} \texttt{CLASS\_EDE} \citep{Hill_2020}, which evolves the EDE field as a pseudo scalar field up to linear order in perturbations; and \texttt{CLASS-PT} \citep{Chudaykin_2020}, which is based on the Effective Field Theory (EFT) of LSS \citep{Baumann_2012, Carrasco_2012} and allows to model the galaxy power spectrum up to mildly nonlinear scales. We perform an MCMC inference with \texttt{MontePython} \citep{Brinckmann_2018}, using the Metropolis--Hastings algorithm \citep{Metropolis_1953, Hastings_1970}. 

Our data set consists of the \textit{Planck} 2018 TT+TE+EE+low$\ell$+lensing likelihoods \citep{Planck_col_2020} along with the BOSS DR 12 full-shape likelihood based on the EFT of LSS presented in \citet{Ivanov_2020_full-shape,DAmico_2020}. Note that this is a slightly different data set than in \citet{Ivanov_2020} and \citet{Smith_2021}, who also included the BOSS (reconstructed) BAO likelihood. We have checked that including the reconstructed BAO data in addition does not lead to a large change of our conclusions. Recently, there has been an update on the BOSS window function from \citet{Beutler:2021eqq} that might impact the conclusions in previous analysis cited here. To compare with the published constraints, we do not use the new window functions. 

We sample the \LCDM parameters $\omega_b$, $\omega_\mathrm{CDM}$, $\theta_s$, $A_s$, $n_s$, $\tau_\mathrm{reio}$ assuming flat priors, along with the \textit{Planck} and EFT nuisance parameters. In Section~\ref{sec:grid}, we assume $f_\EDE \in [0.001, 0.5]$, in Sections~\ref{sec:profile_likelihood},~\ref{sec:Constructing}, we assume $\theta_i \in [0.1,3.1]$ and $\log(z_c) \in [3, 4.3]$. Following the convention of the \textit{Planck} collaboration \citep{Planck_col_2020}, we model the neutrino sector by two massless and one massive neutrino species with $m_\nu = 0.06\,$eV. 

%%%%%%%%%%%%%%%%%%%%%%%%%%%%%%%%%%%

\subsection{Grid Sampling}
\label{sec:grid}

In this section, we perform our first analysis to study how much the conclusions of \citet{Smith_2021} drawn from the 1-parameter model depend on the particular choice of $\theta_i=2.775$ and $\log(z_c)=3.569$. An exploration of the effect of $\theta_i$, $z_c$ on cosmological observables can be found in \cite{Smith_2020, Poulin_2019, Lin_2019}.

\begin{figure}
	\centering
	\includegraphics[scale=.6]{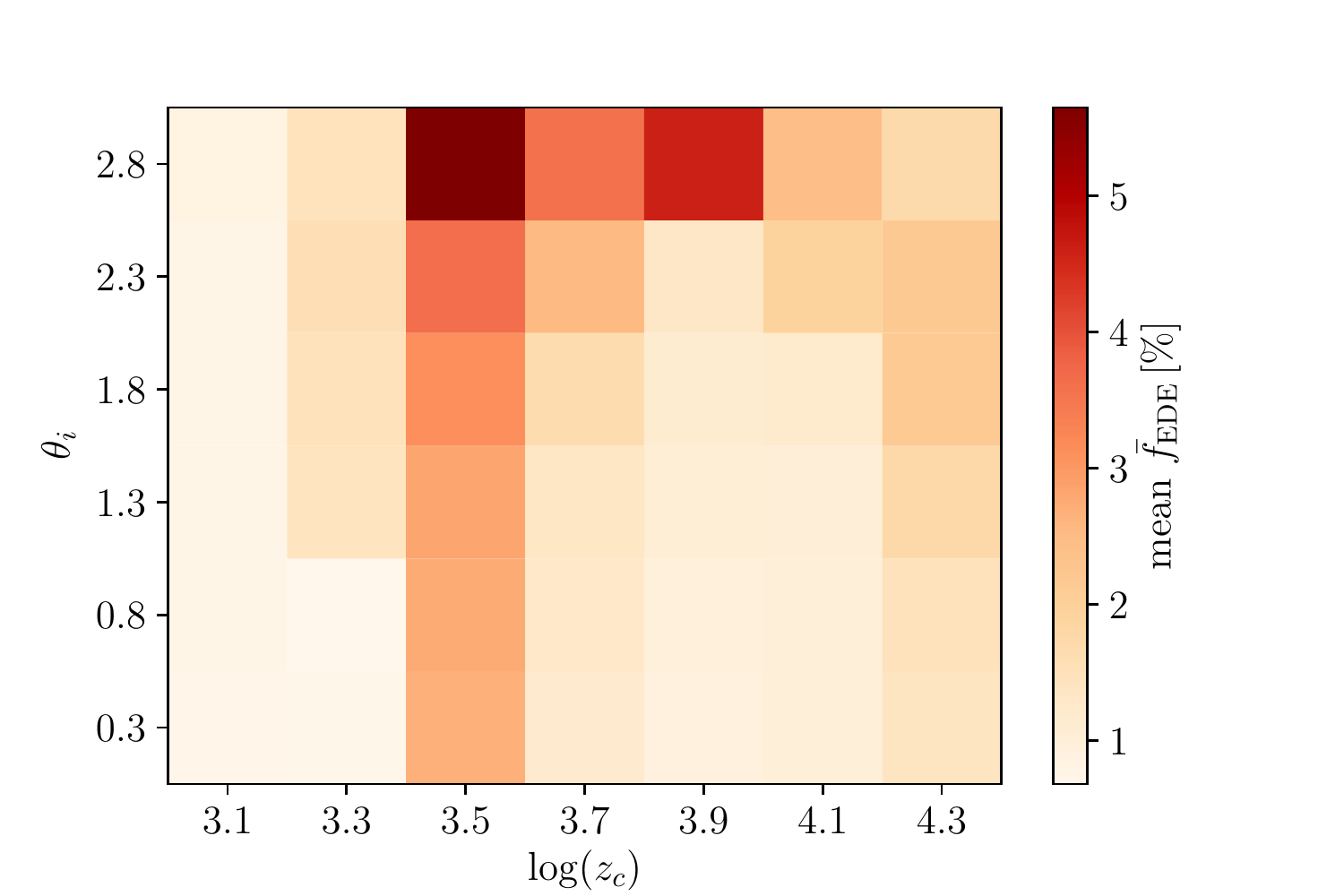}
	\caption{Mean values of $f_\EDE$ for different fixed values of $\theta_i$ and $\log(z_c)$. Every value in this $6\times7$ grid is determined by a full MCMC analysis.}
	\label{fig:grid_mean_fEDE}
\end{figure}

The potential problem encountered in the MCMC exploration of the 3-parameter model is a preference for small $f_\EDE$ 
due to volume effects upon marginalization over $\theta_i$ and $z_c$. We investigate this problem as follows. 
To explore the dependence of the $f_\EDE$ constraints on $\theta_i$ and $z_c$, we run several MCMC inferences, where we keep $\theta_i$ and $z_c$ fixed to different values and vary only $f_\EDE$. We choose six values in the typical prior range of $\theta_i \in [0.1,3.1]$ and seven values in the typical prior range of $\log(z_c) \in [3, 4.3]$:
\begin{align*}
    &\theta_i = \left\{0.3,\, 0.8,\, 1.3,\, 1.8,\, 2.3,\, 2.8 \right\}, \\
    &\log(z_c) = \left\{ 3.1,\, 3.3,\, 3.5,\, 3.7,\, 3.9,\, 4.1,\, 4.3 \right\}.
\end{align*}
Throughout this paper ``$\log$'' denotes the logarithm with base 10. This gives a $6\times7$ grid of MCMC analyses. For each MCMC, we infer the mean fraction of EDE $\bar{f}_\EDE$ depending on the choice of $\theta_i$ and $z_c$. We run every MCMC until the Gelman-Rubin convergence criterion $R-1 < 0.1$ is reached. Our results are summarized in Figure~\ref{fig:grid_mean_fEDE}. 

We find that $\bar{f}_\EDE$ strongly depends on the particular choice of $\theta_i$ and $\log(z_c)$. 
There are choices of $\theta_i$ and $\log(z_c)$ that allow for higher $f_\EDE$. For example, $\theta_i= 2.8$ and $\log(z_c) = 3.5$ (which is close to the values chosen by \citet{Smith_2021}: $\theta_i= 2.775$, $\log(z_c) = 3.569$), allows for particularly high $f_\EDE = 0.057^{+0.027}_{-0.034}$, the highest found in the grid.  The authors point out that this choice of $\left\{ \theta_i, z_c \right\}$ is reasonable since it is obtained from the bestfit cosmology to \textit{Planck} data. However, for $\theta_i= 1.8$ and $\log(z_c) = 3.7$ (which is similar to the mean values found in \citet{Ivanov_2020}: $\theta_i=2.023$, $\log(z_c) = 3.71$), we find $f_\EDE = 0.017^{+0.004}_{-0.016}$. This shows that the particular choice of $\theta_i$ and $\log(z_c)$ made in \citet{Smith_2021} is the reason for a higher $f_\EDE$ than found in \citet{Ivanov_2020}. We point out that the bestfit and mean values quoted in \citet{Ivanov_2020} (bestfit values: $\theta_i=2.734$, $\log(z_c)=3.52$) correspond to choices of $\theta_i$ and $\log(z_c)$ that allow for high and low values of $f_\EDE$, respectively.

We also explore the dependence of the bestfit $f_\EDE$ and the $\Delta\chi^2$ as a function of $\theta_i$ and $z_c$ in Appendix~\ref{app:grid}, finding a similar pattern as in Figure~\ref{fig:grid_mean_fEDE}. We show that the choice of $\theta_i= 2.8$ and $\log(z_c) = 3.5$, which gives the highest mean and bestfit of $f_\EDE$, has the smallest $\chi^2$.

As the constraint on $f_\EDE$ depends strongly on the particular choice of $\left\{ \theta_i, z_c \right\}$, the analysis of the 1-parameter model presented in \citet{Smith_2021} might have been biased. 
Our result also shows that, if $\left\{ \theta_i, z_c \right\}$ cannot be constrained, as in the MCMC analysis of the 3-parameter model \citep{Ivanov_2020}, it might lead to misleading constraints on $f_\EDE$. 

Our grid method is not plagued by volume effects since there is no larger prior volume at $f_\EDE = 0$ compared to $f_\EDE >0$ when $\theta_i$ and $\log(z_c)$ are fixed. This course-grained exploration of the $\left\{ \theta_i, z_c \right\}$ parameter space made with the grid shows that higher values of $f_\EDE$ are allowed for a considerable part of the parameter space and present a good fit to the data. This indicates that volume effects might be present in the 3-parameter MCMC analysis, and that, when this effect is eliminated, the preference for smaller $f_\EDE$ in the posterior is weakened. 

Motivated by this, in the next section we perform a frequentist analysis 
using profile likelihoods, which does not suffer from volume effects.

%%%%%%%%%%%%%%%%%%%%%%%%%%%%%%%%%%%

\subsection{Profile likelihood}
\label{sec:profile_likelihood}

Comparison of the results obtained from Bayesian and frequentist analyses is useful for checking
if priors or marginalization affect the results \citep{Cousins_1995}.
To construct a profile likelihood, one fixes the parameter of interest, i.e. in our case $f_\EDE$, to different values and maximizes the likelihood ${\mathcal L}$ (or minimizes $\chi^2=-2\ln {\mathcal L}$) with respect to all the other parameters of the model, i.e. all $\Lambda$CDM parameters, $\theta_i$ and $z_c$, as well as all the nuisance parameters, for every choice of the parameter of interest ($f_\EDE$). The $\Delta\chi^2$ as a function of the parameter of interest is the profile likelihood \citep[see, e.g.,][for an application to the \textit{Planck} data]{Planck_col_2014}.

For the minimization, we adopt the method used in \citet{Schoeneberg_2021_H0}. For every fixed value of $f_\EDE$, we first run a long MCMC (with at least $10^4$ accepted steps) until the Gelman-Rubin criterion $R-1 < 0.25$ is reached. This yields a reasonable estimate for the bestfit values and covariance of all the other parameters. Second, we run three small chains with successively decreasing step size (decreasing temperature) and enhanced sensitivity to the likelihood difference. This is done with a slightly modified Metropolis--Hastings algorithm as described in \citet{Schoeneberg_2021_H0}.  Since they found that in the context of EDE and other solutions to the Hubble tension, this method was less likely to get stuck in local minima than algorithms based on gradient descent such as MIGRAD \citep{James_1975}, we adopted the same approach.

\begin{figure}
	\centering
	\includegraphics[scale=.6]{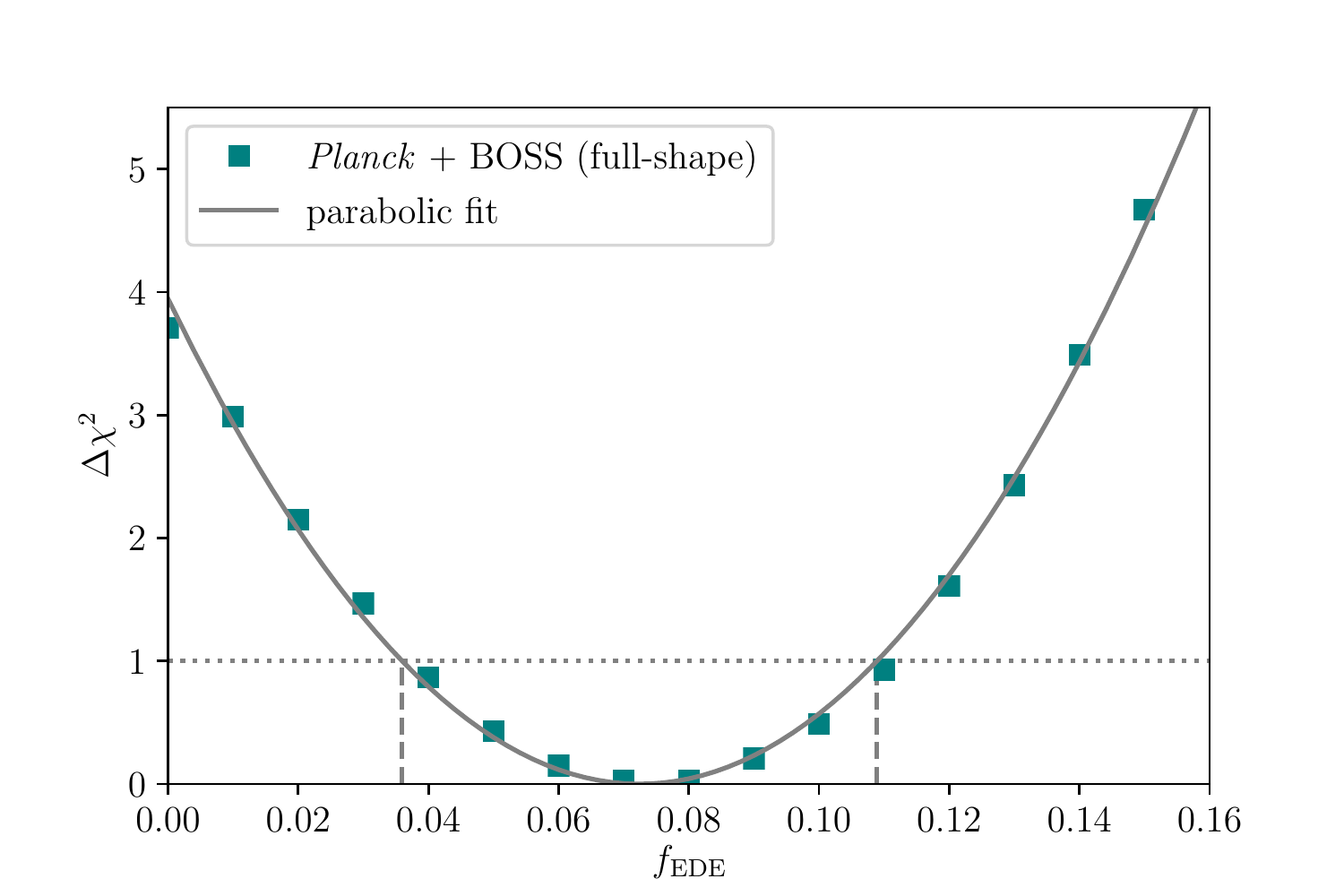}
	\caption{Profile likelihood of the fraction of EDE $f_\EDE$ from the \textit{Planck} CMB and the BOSS full-shape galaxy clustering data. We show $\Delta\chi^2=-2\ln ({\cal L}/{\cal L}_\mathrm{max})$, where ${\cal L}_\mathrm{max}$ is the maximum likelihood, (green markers) and a parabola fit (grey line). 
	The confidence interval is constructed using the Feldman--Cousins prescription \citep{Feldman_1998} (vertical dashed lines). It is indistinguishable from the interval constructed from the intersection of the parabola with $\Delta\chi^2 = 1$ (horizontal dotted line).}
	\label{fig:profile_likelihood_fEDE}
\end{figure}

The results of the minimization are shown as the markers in Figure~\ref{fig:profile_likelihood_fEDE}. For a parameter following a Gaussian distribution, one would expect a parabola, which is 
a good fit for $f_\EDE < 0.15$ (grey line).
The minimum of the curve is the minimum $\chi^2(f_\EDE)$, and shows the bestfit value for $f_\EDE$. Already from the profile likelihood, one can see that our bestfit value lies near the upper bound $f_\EDE < 0.072$ (95\%~C.L) of \cite{Ivanov_2020}. This is a strong indication that the MCMC analysis of the 3-parameter model is plagued by volume effects. The profile likelihood does not suffer from volume effects, since the minimum $\chi^2(f_\EDE)$ is the same as the maximum likelihood estimate.

We report the bestfit values of all the parameters for $f_\EDE=0$, 0.07 and 0.11 in Appendix \ref{app:tables}. We found that the bestfit values of $\left\{ \theta_i, z_c \right\}$ are approximately constant for all fixed values of $f_\EDE$ and fluctuate within a few percent around $\log(z_c) = 3.56$ and $\theta_i = 2.75$. Note that these values are very close to the ones adopted in the 1-parameter model in \citet{Smith_2021}.

%%%%%%%%%%%%%%%%%%%%%%%%%%%%%%%%%%%

\section{Constructing Confidence Intervals: Profile likelihood}
\label{sec:Constructing}

To construct confidence intervals from the profile likelihood shown in Figure~\ref{fig:profile_likelihood_fEDE}, we use the prescription introduced by \citet{Feldman_1998}, which is suitable for a parameter with a physical boundary like $f_\EDE$, which has to lie between 0 and 1. The Feldman--Cousins prescription is based on the likelihood ratio
\begin{equation}
    R(x) = \frac{\mathcal{L}(x|\mu)}{\mathcal{L}(x|\mu_\mathrm{best})},
\end{equation}
where $x$ is the observable or measured value (it can take on all possible values for $f_\EDE$), $\mu$ is the true value of $f_\EDE$ (which will be read off at the minimum of the parabola), and $\mu_\mathrm{best}$ is the physically allowed value $\mu$ for which for a given $x$ the likelihood $\mathcal{L}(x|\mu)$ is maximized; since $\mu_\mathrm{best}>0$, it is $\mu_\mathrm{best}=x$ for $x\geq0$ and $\mu_\mathrm{best}=0$ for $x<0$. The confidence interval $[x_1,\, x_2]$ is chosen such that $R(x_1) = R(x_2)$ and
\begin{equation}
    \int_{x_1}^{x_2} \mathcal{L}(x|\mu)\,\mathrm{d}x = \alpha,
\end{equation}
where $\alpha$ is the confidence level, e.g. $\alpha = 0.6827$ for 68.27\%~C.L. To shorten the notation, we denote 68.27\%~C.L as 68\%~C.L. in the remainder of the paper. For a given $\mu$, the integral is solved numerically and tabulated by \citet{Feldman_1998}. The Feldman--Cousins prescription unambiguously determines whether one parameter should be quoted as an upper/lower limit or as a central confidence interval. Here, we find a central confidence interval at the $68\%$~C.L. By reading off $\mu$ at the minimum of the parabola shown in Figure~\ref{fig:profile_likelihood_fEDE}, we find $f_\EDE = 0.072 \pm 0.036$ ($f_\EDE = 0.072_{-0.060}^{+0.071}$ at $95\%$ C.L.). 

The upper and lower bounds of the $68\%$ confidence interval are shown in Figure~\ref{fig:profile_likelihood_fEDE}, as the vertical dashed lines. They coincide with the confidence intervals constructed by the Neyman prescription \citep{Neyman:1937uhy} (interval between parabola points that intersect with $\Delta\chi^2 =1$) which is only valid far away from a physical boundary.

%%%%%%%%%%%%%%%%%%%%%%%%%%%%%%%%%%%

\section{Discussion and conclusion}
\label{sec:Discussion}

In this paper, we used the grid sampling and profile likelihood methods to understand the difference in the constraints on the EDE model reported in the literature \citep{Ivanov_2020, DAmico_2021,Smith_2021}, using the \textit{Planck} CMB and the BOSS full-shape galaxy clustering data.
With the grid sampling, we showed that the inferred mean and bestfit values of $f_\EDE$ depend strongly on the values of $\left\{ \theta_i, z_c \right\}$. This finding is relevant, since the posterior distributions in the full 3-parameter model shown in \citet{Ivanov_2020} (their Figure~5) indicate that $\theta_i$ and particularly $z_c$ are poorly constrained by the \textit{Planck} and BOSS data. 
Also, depending on the particular choice of $\left\{ \theta_i, z_c \right\}$ made in the 1-parameter model, one could draw different conclusions about the amount of EDE allowed by the data. The choice made in \citet{Smith_2021} is an example of a choice that allows for high value of $f_\EDE$ and therefore a larger effect on $H_0$. However, even for the choice $\theta_i= 2.8$ and $\log(z_c) = 3.5$, which gives the highest value of $f_\EDE$ in our grid method, we find $H_0=69.52_{-1.21}^{+0.95}\,$km/s/Mpc, which only partially alleviates the Hubble tension.

Based on the hints of the grid analysis, we constructed the profile likelihood for $f_\EDE$, which is not subject to volume effects upon marginalization in the MCMC chain. Using the Feldman--Cousins prescription, we constructed the confidence interval, finding $f_\EDE = 0.072 \pm 0.036$, providing a new and robust constraint on the EDE model.

\begin{figure}
	\vspace*{0.5cm}
	\centering
	\includegraphics[scale=.6]{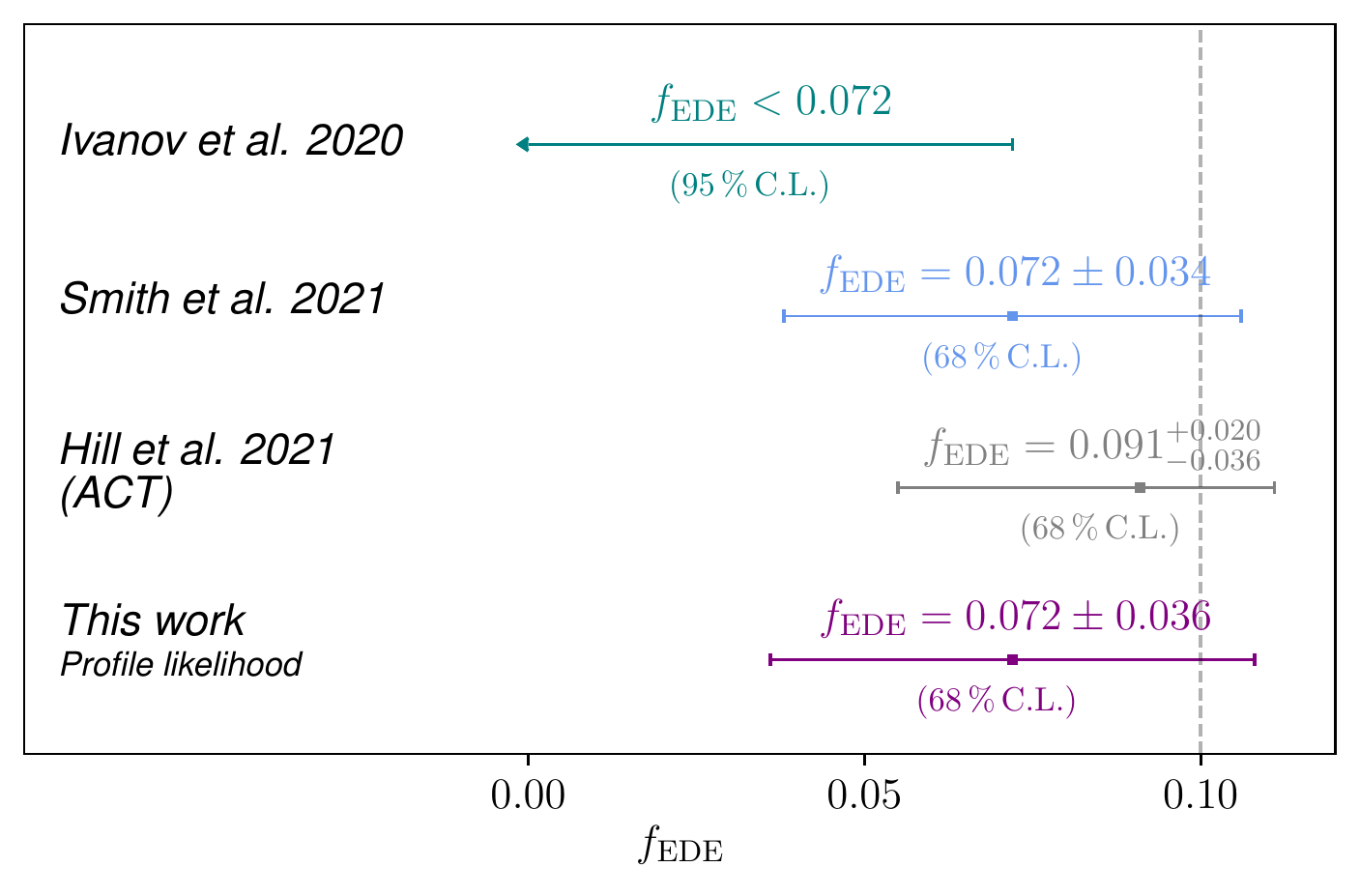}
	\caption{Summary of the current constraints on $f_\EDE$ from the \textit{Planck} CMB and the BOSS full-shape galaxy clustering data by different methods: \citet{Ivanov_2020} with an MCMC inference of the 3-parameter model in green ($95\%$ C.L.), \citet{Smith_2021} with an MCMC inference within the 1-parameter model in blue ($68\%$ C.L.), and our results obtained with the Feldman--Cousins prescription based on the profile likelihood in purple ($68\%$ C.L.). For comparison, we show the recent ACT results in grey \citep{Hill_2021_ACT} ($68\%$ C.L.). The vertical grey dashed line marks $f_\EDE = 0.1$.}
	\label{fig:summary_plot}
\end{figure}

In Figure~\ref{fig:summary_plot}, we compare the confidence interval from this work based on the profile likelihood to previous work. For reference, we mark $f_\EDE = 0.1$. Our bestfit value, $f_\EDE = 0.072$, is at the $95\%$-confidence upper limit found in \citet{Ivanov_2020}, which is $f_\EDE <0.072$. This shows that there is an effect in the MCMC analysis that drives the constraint on $f_\EDE$ closer to zero. The most plausible explanation is volume effects upon marginalization due to the large prior volume in $\theta_i$ and $z_c$ when $f_\EDE \rightarrow 0$.
On the other hand, our bestfit value and the $68\%$~C.L. are similar to those found in \citet{Smith_2021} with the same central value and only slightly larger confidence interval. Nevertheless, their result was obtained within the 1-parameter model, which has a strong dependence on the particular choice of $\theta_i$ and $z_c$ as shown in Section~\ref{sec:grid}, and cannot be used to draw conclusions about the full 3-parameter model.

We suggest that the profile likelihood is a more suitable method to analyze the EDE model, and determine $f_\EDE$. The confidence intervals obtained through this method do not suffer from volume effects or a reduced parameter space. 

We did not construct confidence intervals for $H_0$ in this paper. 
This study is currently in progress together with the analysis of the EDE model with different data sets.

%%%%%%%%%%%%%%%%%%%%%%%%%%%%%%%%%%%

\section*{Acknowledgements}
We thank Paolo Campeti, Elisabeth Krause, Evan McDonough, Marta Monelli, Oliver Philcox, Fabian Schmidt, Sherry Suyu, and Sam Witte for useful discussions and suggestions. We also thank the organizers and participants of the Munich Institute for Astro- and Particle Physics (MIAPP) workshop ``Accelerating Universe 2.0'' for useful discussions about this project during the workshop.
This work was supported in part by JSPS KAKENHI Grant No.~JP20H05850 and JP20H05859, and the Deutsche Forschungsgemeinschaft (DFG, German Research Foundation) under Germany's Excellence Strategy - EXC-2094 - 390783311.
The Kavli IPMU is supported by World Premier International Research Center Initiative (WPI), MEXT, Japan. 

\bibliography{main}{}
\bibliographystyle{aasjournal}

%%%%%%%%%%%%%%%%%%%%%%%%%%%%%%%%%%%

\appendix

%%%%%%%%%%%%%%%%%%%%%%%%%%%%%%%%%%%

\section{Bestfit and \texorpdfstring{$\Delta\chi^2$}{} of grid analysis}
\label{app:grid}

\begin{figure}[h]
    \centering
    \includegraphics[scale=0.55]{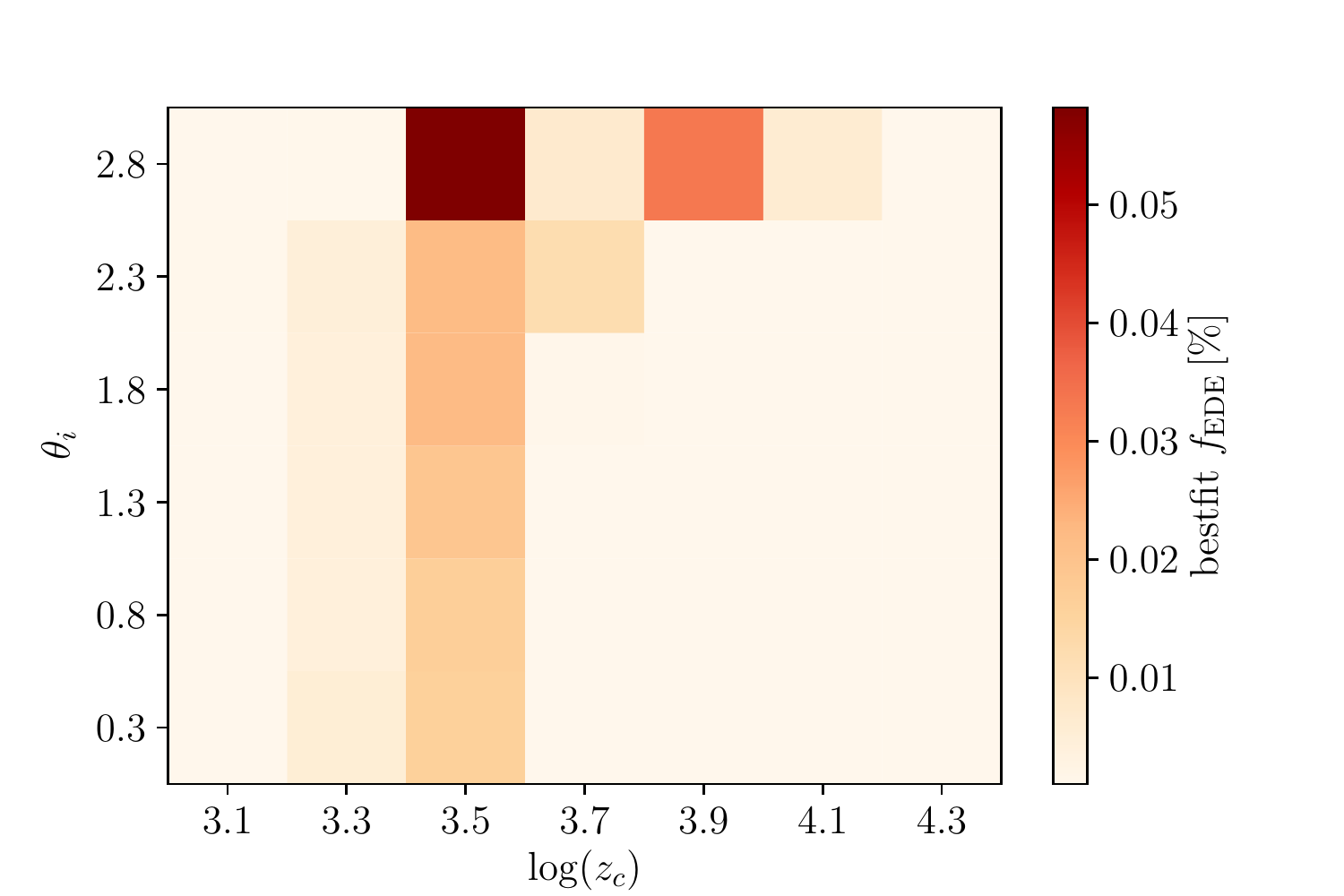} 
    \includegraphics[scale=0.55]{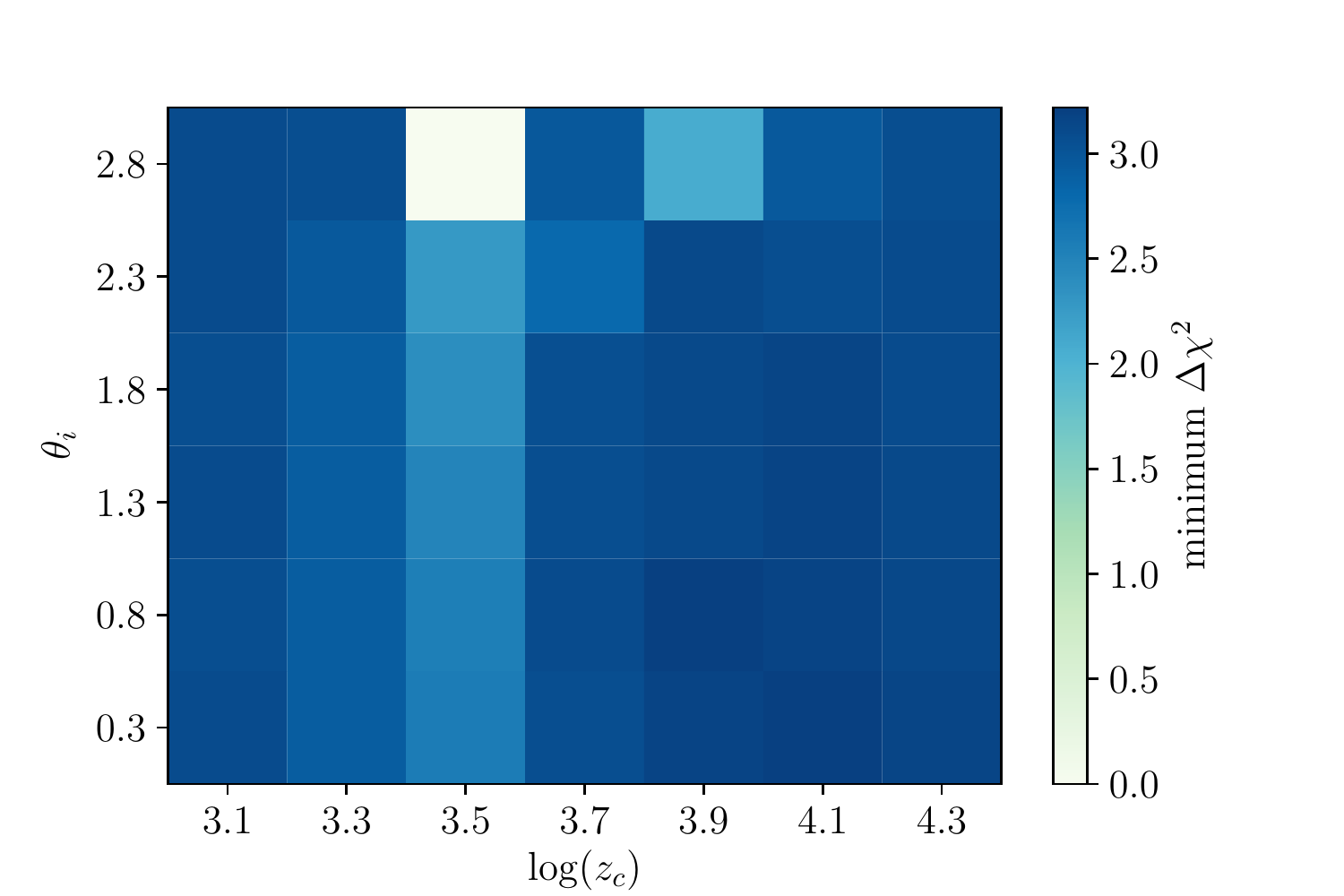}
    \caption{Bestfit values of $f_\EDE$ (\textit{left}) and $\Delta\chi^2$ (\textit{right}) for different fixed values of $\theta_i$ and $\log(z_c)$. Every value in this $6\times7$ grid is obtained with the minimization procedure described in Section~\ref{sec:profile_likelihood}.}
	\label{fig:grid_bestfit_chi2}
\end{figure}

The results of the grid analysis, showing the dependence of the bestfit $f_\EDE$ and the $\Delta\chi^2$ as a function of $\theta_i$ and $z_c$, can be seen in Figure~\ref{fig:grid_bestfit_chi2}.

%%%%%%%%%%%%%%%%%%%%%%%%%%%%%%%%%%%

\section{Bestfit values of the parameters for different cosmologies}
\label{app:tables}

In Table~\ref{tab:bestfits}, we show the bestfit parameters obtained with the minimization described in Section~\ref{sec:profile_likelihood} for the $\Lambda$CDM cosmology and for EDE cosmologies with fixed $f_\EDE$. The first 8 parameters in the table are varied in the MCMC, the last 6 parameters are derived parameters. At the bottom, we quote the minimum $\chi^2$. The cosmology with fixed $f_\EDE=0.07$ is close to the bestfit computed from the minimum of the parabola fit, $f_\EDE=0.11$ is at the higher end of the $68\%$ confidence interval.

%\vspace{0.5cm}
\begin{table}[h]
    \caption{Bestfit parameters for different cosmologies. }
    \centering
    \begin{tabular}{|l|c|c|c|c|}
    \hline
    Parameter & bestfit $\Lambda$CDM & bestfit $f_\EDE = 0.07$ & bestfit $f_\EDE = 0.11$\\ \hline
    $100~\omega_{b }$    &$2.245$  &$2.259$  &$2.270$   \\
    $\omega_\mathrm{cdm}$&$0.1191$ &$0.1260$ &$0.1304$  \\
    $100*\theta_{s }$    &$1.042$  &$1.042$  &$1.041$   \\
    $\ln(10^{10}A_{s})$  &$3.044$  &$3.056$  &$3.064$   \\
    $n_{s}$              &$0.9681$ &$0.9794$ &$0.9872$  \\
    $\tau_\mathrm{reio}$ &$0.0548$ &$0.0549$ &$0.0553$  \\
    $\log(z_{c})$        & --      &$3.55$   &$3.56$    \\
    $\theta_i$           & --      &$2.76$   &$2.77$    \\\hline
    $z_\mathrm{reio}$    &$7.701$  &$7.827$  &$7.924$   \\
    $\Omega_\mathrm{m}$  &$0.3093$ &$0.3046$ &$0.3012$  \\
    $Y_\mathrm{He}$      &$0.2454$ &$0.2479$ &$0.2480$  \\
    $H_0$ [km/s/Mpc]     &$67.80$  &$70.00$  &$71.45$   \\
    $10^{+9}A_{s }$      &$2.099$  &$2.125$  &$2.141$   \\
    $\sigma_8$           &$0.808$  &$0.825$  &$0.836$   \\\hline
    min. $\chi^2$        &$3237.4$ &$3233.7$ &$3234.6$  \\
    \hline
    \end{tabular}
    \label{tab:bestfits}
\end{table}

\end{document}